\documentclass[aps,preprint,nofootinbib,floatfix]{revtex4}%
\usepackage{amssymb}
\usepackage{amsmath}
\usepackage{epsfig}
\usepackage{amsfonts}
\usepackage{graphicx}%
\begin{document}
\title{Stirling number identities and High energy String Scatterings}
\author{Jen-Chi Lee}
\email{jcclee@cc.nctu.edu.tw}
\affiliation{Department of Electrophysics, National Chiao-Tung University and Physics
Division, National Center for Theoretical Sciences, Hsinchu, Taiwan, R.O.C.}
\author{Yi Yang}
\email{yiyang@mail.nctu.edu.tw}
\affiliation{Department of Electrophysics, National Chiao-Tung University and Physics
Division, National Center for Theoretical Sciences, Hsinchu, Taiwan, R.O.C.}
\author{Sheng-Lan Ko}
\email{slko.py96g@g2.nctu.edu.tw}
\affiliation{Department of Electrophysics, National Chiao-Tung University, Hsinchu, Taiwan,
R.O.C. }
\date{\today }

\begin{abstract}
We use Stirling number identities developed recently in number theory to show
that ratios among high energy string scattering amplitudes in the fixed angle
regime can be extracted from the Kummer function of the second kind. This
result not only brings an interesting bridge between string theory and
combinatoric number theory but also sheds light on the understanding of
algebraic structure of high energy stringy symmetry.

\end{abstract}
\maketitle

\section{Introduction}

High energy behaviors of scattering amplitudes are of fundamental importance
in quantum mechanics, quantum field theory and string theory. Not only can
they be used to simplify the mathematical calculation of the amplitudes but
also that one can use the high energy amplitudes to extract many fundamental
characteristics of the physical theory. There are two fundamental regimes of
high energy scattering amplitudes, namely, the fixed angle regime and the
fixed momentum transfer regime. These two regimes represent two different high
energy perturbation expansions of the scattering amplitudes, and contain
complementary information of the underlying theory. In QCD, for example, the
probe of high energy, fixed angle regime reveals the partonic structures of
hadrons, quarks and gluons. On the other hand, the Regge behavior of high
energy hadronic scattering amplitudes suggested a string model of hardons with
a linear relation between hadron spins and their mass squared. In string
theory, the scattering amplitudes in the high energy, fixed angle regime
\cite{GM, Gross, GrossManes}, the Gross regime (GR), were recently
reinvestigated for massive string states at arbitrary mass levels
\cite{ChanLee1,ChanLee2, CHL,CHLTY,PRL,susy,Closed}. The calculations were
carried out by using three different methods, the decoupling of zero-norm
states (ZNS) \cite{ZNS1,ZNS3,ZNS2}, the saddle-point method and the method of
Virasoro constraint. All three methods gave the consistent results. In
particular, an infinite number of linear relations, or stringy symmetries,
among string scattering amplitudes of different string states were obtained.
These linear relations can be solved for each fixed mass level $M^{2}=2(N-1)$,
and ratios $T^{(N,2m,q)}/T^{(N,0,0)},N\geq2m+2q;m,q\geq0$ among the amplitudes
at each fixed mass level can be obtained.

An important question one could ask then is the interpretation of these
infinite number of ratios in terms of the concept of symmetry. For example,
does there exist any algebraic structure (or group structure) of these ratios?
Since so far little has been known for the full spacetime symmetry\ of 26D
string theory (except $\omega_{\infty}$ for the case of toy 2D string theory
\cite{ZNS3}), mathematically the meaning of these infinite number of ratios
remains mysterious. One of our recent research is to understand this issue
from various directions. To our surprise, it turns out that some of the
questions above can be answered by calculating high energy string scatterings
in another regime, the fixed momentum transfer regime, or the Regge regime (RR).

\bigskip

There are two main results of this report. First, we discover that the leading
order amplitudes at each fixed mass level in the RR can be expressed in terms
of the Kummer function of the second kind. Second, the number of high energy
scattering amplitudes for each fixed mass level in the RR is much more
numerous than that of GR. For those leading order high energy amplitudes
$A^{(N,2m,q)}$ in the RR with the same type of $(N,2m,q)$ as those of GR, we
can extract from them the above mentioned ratios $T^{(N,2m,q)}/T^{(N,0,0)}$ in
the GR by using Kummer function of the second kind, which naturally shows up
in the leading order of high energy string scattering amplitudes in the RR.
The calculation brings a link between high energy string scattering amplitudes
in the GR and the RR. On the other hand, it seems that both the saddle-point
method and the method of decoupling of high energy ZNS adopted in the
calculation of GR do not apply to the case of RR, and there is no linear
relation anymore as in the case of scatterings in the GR. For more results on
high energy string scatterings in the RR, see \cite{RR1,RR2,RR3,RR4,RR5,RR6}.

We stress that, mathematically, the proof of the identification of the ratios
in the GR from the Kummer function calculated in the RR turns out to be highly
nontrivial. This is based on a summation algorithm for Stirling number
identity derived by Mkauers in 2007 \cite{MK}. It is very interesting to see
that the identity in Eq.(\ref{14}) suggested by string theory calculation can
be rigorously proved by a totally different mathematical method. Although this
kind of coincidence is not unusual in the development of string theory, our
results bring an interesting connection between string theory and combinatoric
number theory. Moreover, the connection between Kummer function and high
energy string scatterings may shed light on a deeper understanding of stringy symmetries.

\section{\bigskip Regge Scatterings}

We begin with a brief review of high energy string scatterings in the GR. That
is in the kinematic regime%
\begin{equation}
s,-t\rightarrow\infty,t/s\approx-\sin^{2}\frac{\theta}{2}=\text{fixed (but
}\theta\neq0\text{)}%
\end{equation}
where $s,t$ and $u$ are the Mandelstam variables and $\theta$ is the CM
scattering angle. It was shown \cite{CHLTY,PRL} that for the 26D open bosonic
string the only states that will survive the high-energy limit at mass level
$M_{2}^{2}=2(N-1)$ are of the form
\begin{equation}
\left\vert N,2m,q\right\rangle \equiv(\alpha_{-1}^{T})^{N-2m-2q}(\alpha
_{-1}^{L})^{2m}(\alpha_{-2}^{L})^{q}\left\vert 0,k_{2}\right\rangle \label{1}%
\end{equation}
where the polarizations of the 2nd particle with momentum $k_{2}$ on the
scattering plane were defined to be $e^{P}=\frac{1}{M_{2}}(E_{2}%
,\mathrm{k}_{2},0)=\frac{k_{2}}{M_{2}}$ as the momentum polarization,
$e^{L}=\frac{1}{M_{2}}(\mathrm{k}_{2},E_{2},0)$ the longitudinal polarization
and $e^{T}=(0,0,1)$ the transverse polarization. Note that $e^{P}$ approaches
to $e^{L}$ in the GR, and the scattering plane is defined by the spatial
components of $e^{L}$ and $e^{T}$. Polarizations perpendicular to the
scattering plane are ignored because they are kinematically suppressed for
four point scatterings in the high-energy limit. One can then use the
saddle-point method to calculate the high energy scattering amplitudes. For
simplicity, we choose $k_{1}$, $k_{3}$ and $k_{4}$ to be tachyons and the
final result of the ratios of high energy, fixed angle string scattering
amplitude are \cite{CHLTY,PRL}%
\begin{equation}
\frac{T^{(N,2m,q)}}{T^{(N,0,0)}}=\left(  -\frac{1}{M_{2}}\right)
^{2m+q}\left(  \frac{1}{2}\right)  ^{m+q}(2m-1)!!.\label{2}%
\end{equation}
The ratios in Eq.(\ref{2}) can also be obtained by using the decoupling of two
types of ZNS in the spectrum. As an example, for $M_{2}^{2}=4$ we get
\cite{ChanLee1,ChanLee2}
\begin{equation}
T_{TTT}:T_{LLT}:T_{(LT)}:T_{[LT]}=8:1:-1:-1.\label{3}%
\end{equation}
To convince the readers that the infinite ratios in Eq.(\ref{2}) are the
symmetries or, at least, remnant of full spacetime symmetries of 26D string
theory, it was shown that a set of 2D discrete ZNS $\Omega_{J_{1},M_{1}}^{+}$
carry $\omega_{\infty}$ symmetry charges \cite{ZNS3}%
\begin{equation}
\int\frac{dz}{2\pi i}\Omega_{J_{1},M_{1}}^{+}(z)\cdot\Omega_{J_{2},M_{2}}%
^{+}(0)=(J_{2}M_{1}-J_{1}M_{2})\Omega_{(J_{1}+J_{2}-1),(M_{1}+M_{2})}%
^{+}(0).\label{26}%
\end{equation}

A natural question arises. Is there any mathematical structure (e.g. group
structure) of these infinite number of ratios? Let's consider a simple analogy
from partical physics. The ratios of the nucleon-nucleon scattering processes%
\begin{align}
(a)\text{ \ }p+p  &  \rightarrow d+\pi^{+},\nonumber\\
(b)\text{ \ }p+n  &  \rightarrow d+\pi^{0},\nonumber\\
(c)\text{ \ }n+n  &  \rightarrow d+\pi^{-}%
\end{align}
can be calculated to be%
\begin{equation}
T_{a}:T_{b}:T_{c}=1:\frac{1}{\sqrt{2}}:1 \label{27}%
\end{equation}
from $SU(2)$ isospin symmetry. Similarly, as we will see in this report, the
ratios in Eq.(\ref{2}) can be extracted from Kummer function. The key is to
study high energy string scatterings in the RR.

We now turn to the discussion on high energy string scatterings in the RR.
That is in the kinematic regime%
\begin{equation}
s\rightarrow\infty,\sqrt{-t}=\text{fixed (but }\sqrt{-t}\neq\infty).
\end{equation}
The relevant kinematics in the RR are%
\begin{equation}
e^{P}\cdot k_{1}\simeq-\frac{s}{2M_{2}},\text{ \ }e^{P}\cdot k_{3}\simeq
-\frac{\tilde{t}}{2M_{2}}=-\frac{t-M_{2}^{2}-M_{3}^{2}}{2M_{2}};
\end{equation}%
\begin{equation}
e^{L}\cdot k_{1}\simeq-\frac{s}{2M_{2}},\text{ \ }e^{L}\cdot k_{3}\simeq
-\frac{\tilde{t}^{\prime}}{2M_{2}}=-\frac{t+M_{2}^{2}-M_{3}^{2}}{2M_{2}};
\end{equation}
and%
\begin{equation}
e^{T}\cdot k_{1}=0\text{, \ \ }e^{T}\cdot k_{3}\simeq-\sqrt{-{t}}.
\end{equation}
Note that, unlike the case of GR, $e^{P}$ \textit{does not} approach to
$e^{L}$ in the RR. On the other hand, instead of states in Eq.(\ref{1}) for
the GR, one can argue that the most general string states one needs to
consider at each fixed mass level $N=\sum_{n,m}nk_{n}+mq_{m}$ for the RR are%
\begin{equation}
\left\vert k_{n},q_{m}\right\rangle =\prod_{n>0}(\alpha_{-n}^{T})^{k_{n}}%
\prod_{m>0}(\alpha_{-m}^{L})^{q_{m}}|0\rangle.
\end{equation}
It seems that both the saddle-point method and the method of decoupling of
high energy ZNS adopted in the calculation of GR do not apply to the case of
RR. However the calculation is still manageable, and the general formula for
the high energy scattering amplitudes in the RR can be written down
explicitly. The $s-t$ channel scattering amplitudes of this state with three
other tachyonic states can be calculated to be
\begin{align}
A^{(k_{n},q_{m})}  &  =\int_{0}^{1}dx\,x^{k_{1}\cdot k_{2}}(1-x)^{k_{2}\cdot
k_{3}}\left[  \frac{ie^{L}\cdot k_{1}}{-x}+\frac{ie^{L}\cdot k_{3}}%
{1-x}\right]  ^{q_{1}}\nonumber\\
&  \cdot\prod_{n=1}\left[  \frac{ie^{T}\cdot k_{3}\,(n-1)!}{(1-x)^{n}}\right]
^{k_{n}}\prod_{m=2}\left[  \frac{ie^{L}\cdot k_{3}\,(m-1)!}{(1-x)^{m}}\right]
^{q_{m}}\nonumber\\
&  =\left(  \frac{-i\tilde{t}^{\prime}}{2M_{2}}\right)  ^{q_{1}}\sum
_{j=0}^{q_{1}}{\binom{q_{1}}{j}}\left(  \frac{s}{-\tilde{t}}\right)  ^{j}%
\int_{0}^{1}dxx^{k_{1}\cdot k_{2}-j}(1-x)^{k_{2}\cdot k_{3}+j-\sum
_{n,m}(nk_{n}+mq_{m})}\nonumber\\
&  \cdot\prod_{n=1}\left[  i\sqrt{-t}(n-1)!\right]  ^{k_{n}}\prod_{m=2}\left[
i\tilde{t}^{\prime}(m-1)!\left(  -\frac{1}{2M_{2}}\right)  \right]  ^{q_{m}%
}\nonumber\\
&  =\left(  \frac{-i\tilde{t}^{\prime}}{2M_{2}}\right)  ^{q_{1}}\sum
_{j=0}^{q_{1}}{\binom{q_{1}}{j}}\left(  \frac{s}{-\tilde{t}}\right)
^{j}B\left(  k_{1}\cdot k_{2}-j+1\,,\,k_{2}\cdot k_{3}+j-N+1\right)
\nonumber\\
&  \cdot\prod_{n=1}\left[  i\sqrt{-t}(n-1)!\right]  ^{k_{n}}\prod_{m=2}\left[
i\tilde{t}^{\prime}(m-1)!\left(  -\frac{1}{2M_{2}}\right)  \right]  ^{q_{m}}.
\end{align}
The Beta function above can be approximated in the large $s$, but fixed $t$
limit as follows
\begin{align}
&  B\left(  k_{1}\cdot k_{2}-j+1,k_{2}\cdot k_{3}+j-N+1\right) \nonumber\\
&  =B\left(  -1-\frac{s}{2}+N-j,-1-\frac{t}{2}+j\right) \nonumber\\
&  =\frac{\Gamma(-1-\frac{s}{2}+N-j)\Gamma(-1-\frac{t}{2}+j)}{\Gamma(\frac
{u}{2}+2)}\nonumber\\
&  \approx B\left(  -1-\frac{1}{2}s,-1-\frac{t}{2}\right)  \left(  -1-\frac
{s}{2}\right)  ^{N-j}\left(  \frac{u}{2}+2\right)  ^{-N}\left(  -1-\frac{t}%
{2}\right)  _{j}\nonumber\\
&  \approx B\left(  -1-\frac{1}{2}s,-1-\frac{t}{2}\right)  \left(  -\frac
{s}{2}\right)  ^{-j}\left(  -1-\frac{t}{2}\right)  _{j}%
\end{align}
where%
\begin{equation}
(a)_{j}=a(a+1)(a+2)...(a+j-1)
\end{equation}
is the Pochhammer symbol. The leading order amplitude in the RR can then be
written as%
\begin{align}
A^{(k_{n},q_{m})}  &  =\left(  \frac{-i\tilde{t}^{\prime}}{2M_{2}}\right)
^{q_{1}}B\left(  -1-\frac{1}{2}s,-1-\frac{t}{2}\right)  \sum_{j=0}^{q_{1}%
}{\binom{q_{1}}{j}}\left(  \frac{2}{\tilde{t}^{\prime}}\right)  ^{j}\left(
-1-\frac{t}{2}\right)  _{j}\nonumber\\
&  \cdot\prod_{n=1}\left[  i\sqrt{-t}(n-1)!\right]  ^{k_{n}}\prod_{m=2}\left[
i\tilde{t}^{\prime}(m-1)!\left(  -\frac{1}{2M_{2}}\right)  \right]  ^{q_{m}},
\label{A}%
\end{align}
which is UV power-law behaved as expected. The summation in eq. (\ref{A}) can
be represented by the Kummer function of the second kind $U$ as follows,
\begin{equation}
\sum_{j=0}^{p}{\binom{p}{j}}\left(  \frac{2}{\tilde{t}^{\prime}}\right)
^{j}\left(  -1-\frac{t}{2}\right)  _{j}=2^{p}(\tilde{t}^{\prime}%
)^{-p}\ U\left(  -p,\frac{t}{2}+2-p,\frac{\tilde{t}^{\prime}}{2}\right)  .
\label{equality}%
\end{equation}
Finally, the amplitudes can be written as
\begin{align}
A^{(k_{n},q_{m})}  &  =\left(  -\frac{i}{M_{2}}\right)  ^{q_{1}}U\left(
-q_{1},\frac{t}{2}+2-q_{1},\frac{\tilde{t}^{\prime}}{2}\right)  B\left(
-1-\frac{s}{2},-1-\frac{t}{2}\right) \nonumber\\
&  \cdot\prod_{n=1}\left[  i\sqrt{-t}(n-1)!\right]  ^{k_{n}}\prod_{m=2}\left[
i\tilde{t}^{\prime}(m-1)!\left(  -\frac{1}{2M_{2}}\right)  \right]  ^{q_{m}}.
\label{general amplitude}%
\end{align}
In the above, $U$ is the Kummer function of the second kind and is defined to
be%
\begin{equation}
U(a,c,x)=\frac{\pi}{\sin\pi c}\left[  \frac{M(a,c,x)}{(a-c)!(c-1)!}%
-\frac{x^{1-c}M(a+1-c,2-c,x)}{(a-1)!(1-c)!}\right]  \text{ \ }(c\neq2,3,4...)
\end{equation}
where $M(a,c,x)=\sum_{j=0}^{\infty}\frac{(a)_{j}}{(c)_{j}}\frac{x^{j}}{j!}$ is
the Kummer function of the first kind. $U$ and $M$ are the two solutions of
the Kummer Equation%
\begin{equation}
xy^{^{\prime\prime}}(x)+(c-x)y^{\prime}(x)-ay(x)=0.
\end{equation}
It is crucial to note that $c=\frac{t}{2}+2-q_{1},$ and is not a constant as
in the usual case, so $U$ in Eq.(\ref{general amplitude}) is not a solution of
the Kummer equation. This will make our analysis in the next section more
complicated as we will see soon. On the contrary, since $a=-q_{1}$ is an
integer, the Kummer function in Eq.(\ref{equality}) terminated to be a finite
sum. This will simplify the manipulation of Kummer function used in this report.

\section{\bigskip Applying Stirling number identities}

As an important application of Eq.(\ref{general amplitude}), the leading order
amplitudes of string states in the RR, which share the same structure as
Eq.(\ref{1}) in the GR can be written as%
\begin{align}
A^{(N,2m,q)}  &  =B\left(  -1-\frac{s}{2},-1-\frac{t}{2}\right)  \sqrt
{-t}^{N-2m-2q}\left(  \frac{1}{2M_{2}}\right)  ^{2m+q}\nonumber\\
&  2^{2m}(\tilde{t}^{\prime})^{q}U\left(  -2m\,,\,\frac{t}{2}+2-2m\,,\,\frac
{\tilde{t}^{\prime}}{2}\right)  . \label{9}%
\end{align}
It is important to note that there is no linear relation among high energy
string scattering amplitudes of different string states for each fixed mass
level in the RR as can be seen from Eq.(\ref{9}). This is very different from
the result in the GR in Eq.(\ref{2}). In other words, the ratios
$A^{(N,2m,q)}/A^{(N,0,0)}$ are $t$-dependent functions. In particular, we can
extract the coefficients of the highest power of $t$ in $A^{(N,2m,q)}%
/A^{(N,0,0)}$. We can use the identity of the Kummer function%
\begin{align}
&  2^{2m}(\tilde{t}^{\prime})^{-2m}\ U\left(  -2m,\frac{t}{2}+2-2m,\frac
{\tilde{t}^{\prime}}{2}\right) \nonumber\\
&  =\,_{2}F_{0}\left(  -2m,-1-\frac{t}{2},-\frac{2}{\tilde{t}^{\prime}}\right)
\nonumber\\
&  \equiv\sum_{j=0}^{2m}\left(  -2m\right)  _{j}\left(  -1-\frac{t}{2}\right)
_{j}\frac{\left(  -\frac{2}{\tilde{t}^{\prime}}\right)  ^{j}}{j!}\label{12}\\
&  =\sum_{j=0}^{2m}{\binom{2m}{j}}\left(  -1-\frac{t}{2}\right)  _{j}\left(
\frac{2}{\tilde{t}^{\prime}}\right)  ^{j}\nonumber
\end{align}
to calculate
\begin{equation}
\frac{A^{(N,2m,q)}}{A^{(N,0,0)}}=(-1)^{q}\left(  \frac{1}{2M_{2}}\right)
^{2m+q}(-t)^{m}\sum_{j=0}^{2m}(-2m)_{j}\left(  -1-\frac{t}{2}\right)
_{j}\frac{(-2/t)^{j}}{j!}+\mathit{O}\left\{  \left(  \frac{1}{t}\right)
^{m+1}\right\}  \label{13}%
\end{equation}
where we have replaced $\tilde{t}^{\prime}$ by $t$ as $t$ is large. If the
leading order coefficients in Eq.(\ref{13}) extracted from the high energy
string scattering amplitudes in the RR are to be identified with the ratios
calculated previously among high energy string scattering amplitudes in the GR
in Eq.(\ref{2}), we need the following identity
\begin{align}
&  \sum_{j=0}^{2m}(-2m)_{j}\left(  -1-\frac{t}{2}\right)  _{j}\frac
{(-2/t)^{j}}{j!}\nonumber\\
&  =0(-t)^{0}+0(-t)^{-1}+...+0(-t)^{-m+1}+\frac{(2m)!}{m!}(-t)^{-m}%
+\mathit{O}\left\{  \left(  \frac{1}{t}\right)  ^{m+1}\right\}  . \label{14}%
\end{align}
The coefficient of the term $\mathit{O}\left\{  \left(  1/t\right)
^{m+1}\right\}  $ in Eq.(\ref{14}) is irrelevant for our discussion. The proof
of Eq.(\ref{14}) turns out to be nontrivial. The standard approach by using
integral representation of the Kummer function seems not applicable here.
Presumably, the difficulty of the rigorous proof of Eq.(\ref{14}) is
associated with the unusual non-constant $c$ in the argument of Kummer
function in Eqs.(\ref{9}) and (\ref{12}) as mentioned above. It is a
nontrivial task to do the proof compared to the usual cases where the argument
$c$ of the Kummer function is a constant. Here we will adopt another approach
to prove Eq.(\ref{14}). This approach strongly relies on the algorithm for
Stirling number identity derived by Mkauers \cite{MK} in 2007, and is highly
nontrivial either. The leading order identity of Eq.(\ref{14}) can be written
as
\begin{equation}
f(m)\equiv\sum_{j=0}^{m}(-1)^{j}{\binom{2m}{j+m}}\left[
s(j+m-1,j-1)+s(j+m-1,j)\right]  =(2m-1)!! \label{15}%
\end{equation}
where the signed first Stirling number $s(n,k)$ is defined to be
\begin{equation}
(x)_{n}=\sum_{k=0}^{n}(-1)^{n-k}s(n,k)x^{k}. \label{16}%
\end{equation}
The authors had verified the validity of Eq.(\ref{15}) for $m=1,2...,2000$
before they carried out the exact proof to be discussed below. To prove
Eq.(\ref{15}) we define
\begin{equation}
f(u,m)\equiv\sum_{j=0}^{m+u}(-1)^{j}{\binom{2m+u}{j+m}}\left[
s(j+m-1,j-1)+s(j+m-1,j)\right]  \label{17}%
\end{equation}
with $f(0,m)=f(m)$. By using the result of \cite{MK}, one can prove that
$f(u,m)$ satisfies the following recurrence relation
\begin{equation}
-(1+2m+u)f(u,m)+(2m+u)f(u+1,m)+f(u,m+1)=0. \label{18}%
\end{equation}
Eq.(\ref{18}) is the most nontrivial step to prove Eq.(\ref{15}). Finally by
taking $u=0$, it can be shown that the second term of Eq.(\ref{18}) vanishes
\cite{KLY}. Eq.(\ref{15}) is then proved by mathematical induction. The
vanishing of the coefficients of $(-t)^{0},(-t)^{-1},...(-t)^{-m+1}$ terms on
the LHS of Eq.(\ref{14}) means, for $1\leqslant i\leqslant m$,
\begin{equation}
g(m,i)\equiv\sum_{j=0}^{m+i}(-1)^{j-i}{\binom{2m}{j+m-i}}\left[
s(j+m-1-i,j)+s(j+m-1-i,j-1)\right]  =0. \label{19}%
\end{equation}
To prove this identity, we need the recurrence relation \cite{MK}
\begin{align}
-  &  2(1+m)^{2}(1+2m)g(m,i)+(2+7m+4m^{2})g(m+1,i)\nonumber\\
-  &  2m(1+m)(1+2m)g(m+1,i+1)-m\times g(m+2,i)=0. \label{20}%
\end{align}
Putting $i=0,1,2..$, and using the fact we have just proved, i.e.
$g(m+1,0)=(2m+1)g(m,0)$, one can prove Eq.(\ref{19}). Eq.(\ref{14}) is thus
finally proved. It is very interesting to see that the identity in
Eq.(\ref{14}) suggested by string scattering amplitude calculation can be
rigorously proved by a totally different but sophisticated mathematical
method. In conclusion, ratios in Eq.(\ref{2}) can be extracted from Kummer
function of the second kind%
\begin{equation}
\frac{T^{(N,2m,q)}}{T^{(N,0,0)}}=\left(  -\frac{1}{2M}\right)  ^{2m+q}%
2^{2m}\lim_{t\rightarrow\infty}(-t)^{-m}U\left(  -2m\,,\,\frac{t}%
{2}+2-2m\,,\,\frac{t}{2}\right)  . \label{21}%
\end{equation}
In view of Eq.(\ref{27}), this result may help to uncover the fundamental
symmetry of string theory. At last, we give an explicit calculation of the
high energy string scattering amplitudes to subleading orders in the RR for
$M_{2}^{2}=4$ \cite{KLY}
\begin{equation}
A_{TTT}\sim\frac{1}{8}\sqrt{-t}ts^{3}+\frac{3}{16}\sqrt{-t}t(t+6)s^{2}%
+\frac{3t^{3}+84t^{2}-68t-864}{64}\sqrt{-t}\,s+O(1), \label{22}%
\end{equation}%
\begin{align}
A_{LLT}  &  \sim\frac{1}{64}\sqrt{-t}(t-6)s^{3}+\frac{3}{128}\sqrt{-t}%
(t^{2}-20t-12)s^{2}\nonumber\\
&  \quad\quad+\frac{3t^{3}-342t^{2}-92t+5016+1728(-t)^{-1/2}}{512}\sqrt
{-t}\,s+O(1), \label{23}%
\end{align}%
\begin{align}
A_{(LT)}  &  \sim-\frac{1}{64}\sqrt{-t}(t+10)s^{3}-\frac{1}{128}\sqrt
{-t}(3t^{2}+52t+60)s^{2}\nonumber\\
&  \quad\quad-\frac{3[t^{3}+30t^{2}+76t-1080-960(-t)^{-1/2}]}{512}\sqrt
{-t}\,s+O(1), \label{24}%
\end{align}%
\begin{align}
A_{[LT]}  &  \sim-\frac{1}{64}\sqrt{-t}(t+2)s^{3}-\frac{3}{128}\sqrt
{-t}(t+2)^{2}s^{2}\nonumber\\
&  \quad\quad-\frac{(3t-8)(t+6)^{2}[1-2(-t)^{-1/2}]}{512}\sqrt{-t}\,s+O(1).
\label{25}%
\end{align}
We have ignored an overall irrelevant factors in the above amplitudes. Note
that the calculation of Eq.(\ref{24}) and Eq.(\ref{25}) involves amplitude of
the state $(\alpha_{-2}^{T})(\alpha_{-1}^{L})\left\vert 0,k_{2}\right\rangle $
which can be shown to be of leading order in the RR \cite{KLY}, but is of
subleading order in the GR as it is not in the form of Eq.(\ref{1}). However,
the contribution of the amplitude calculated from this state will not affect
the ratios $8:1:-1:-1$ in the RR \cite{KLY}. One can now easily see that the
ratios of the coefficients of the highest power of $t$ in these leading order
coefficient functions $\frac{1}{8}:\frac{1}{64}:-\frac{1}{64}:-\frac{1}{64}$
agree with the ratios in the GR calculated in Eq.(\ref{3}) as expected.
Moreover, one further obeservation is that these ratios remain the same for
the coefficients of the highest power of $t$ in the subleading orders
$(s^{2})$ $\frac{3}{16}:\frac{3}{128}:-\frac{3}{128}:-\frac{3}{128}$ and $(s)$
$\frac{3}{64}:\frac{3}{512}:-\frac{3}{512}:-\frac{3}{512}$. More examples will
be given in \cite{KLY}. We thus conjecture that the existence of these GR
ratios of Eq.(\ref{2}) in the RR persists to the subleading orders in the
Regge expansion of high energy string scattering amplitudes.

\section{Conclusion}

In conclusion, physically, the connection between Kummer function and high
energy string scattering amplitudes derived in this report may shed light on a
deeper understanding of stringy symmetries. Mathematically, the proof of
identity in Eq.(\ref{14}) brings an interesting bridge between string theory
and combinatoric number theory.

\section{Acknowledgement}

JC would like to thank the organizers of 10th workshop on QCD for inviting him
to present this work. This work is supported in part by the National Science
Council, 50 billions project of Ministry of Education and National Center for
Theoretical Science, Taiwan. We appreciated the correspondence of Dr. Manuel
Mkauers at RISC, Austria for his kind help of providing us with the rigorous
proof of Eq.(\ref{15}).

\end{document}